\documentclass[%
preprint,
superscriptaddress,
amsmath,amssymb,
aps,
]{revtex4-1}

\usepackage{graphicx}
\usepackage{dcolumn}
\usepackage{bm}
\usepackage{color}
\usepackage{hyperref}

\def\vec{\mathbf}
\def\DP{^{\prime\prime}}
\def\P{^{\prime}}
\def\P{^{\prime}}
\def\eps{\varepsilon}

\begin{document}

\title{Polarizability of Radially Inhomogeneous Subwavelength Spheres}

\author{Dimitrios C. Tzarouchis}
\email{dimitrios.tzarouchis@aalto.fi}
\affiliation{%
 Department of Electronics and Nanoengineering, Aalto University \\
 Maarintie 8, 02150, Espoo, Finland
}
\affiliation{%
Electrical and Systems Engineering Department, University of Pennsylvania,\\
200 South and 33rd Street, PA 19104-6314, Philadelphia, U.S.A.
}
\author{Ari Sihvola}%
\email{ari.sihvola@aalto.fi}
\affiliation{%
 Department of Electronics and Nanoengineering, Aalto University \\
 Maarintie 8, 02150, Espoo, Finland
}

%
%
%

\date{\today}

\begin{abstract}
In this work the polarizability of a subwavelength core-shell sphere is
considered, where the shell exhibits a radially inhomogeneous permittivity
profile. A mathematical treatment of the electrostatic
polarizability is formulated in terms of the scattering potentials and the
corresponding scattering amplitudes. As a result, a
generalized expression of the polarizability is presented as a function of
the radial inhomogeneity function. The extracted general model is applied
for two particular cases, i.e., the well-known power-law profile and a new class of
permittivity profiles that exhibit exponential radial dependence.
The proposed analysis quantifies in a simple manner the inhomogeneity effects,
allowing the direct implementation of naturally or
artificially occurring permittivity inhomogeneities for a wide range of applications
within and beyond the metamaterial
paradigm. Furthermore, the described analysis open avenues towards the phenomenological
and first-principles modeling of the electrodynamic scattering effects for
graded-index plasmonic particles at the nanoscale. Finally, such description can
be readily used either for the benchmarking of novel computational methods incorporating
inhomogeneous materials or for inverse scattering purposes.

\end{abstract}

\pacs{Valid PACS appear here}
\maketitle

Electromagnetic scattering by subwavelength sphere is a canonical and
fundamental problem found in the core of many areas such as RF/optical
engineering, bioengineering, and material
sciences~\cite{Shore2015,Stewart2008,Krasnok2018}. The significance of this
problem reaches from conventional applications, such as sensing~\cite{Stewart2008}
and energy control/harvesting~\cite{atwater2010plasmonics}, towards more exotic ones,
such as invisibility cloaks~\cite{Alu2005}, super-scatterers~\cite{Ruan2010},
and optical energy localization~\cite{Silveirinha2014}. Studies on
subwavelength spherical scatterers unveiled several fundamental aspects of the
resonant scattering of a sphere, a testbed for extracting physical intuition for
a plethora of phenomena in physics like the plasmon hybridization and Fano
resonances on single scatterers~\cite{Prodan2003,Lukyanchuk2010,Fan2014}.

This universality emerges, perhaps, from the fact that the scattering response
of a small dielectric sphere can be rigorously quantified by a (normalized)
polarizability expression~~\cite{bohren2008absorption}
\begin{equation}
\alpha=3\frac{\eps_1-\eps_h}{\eps_1+2\eps_h}
\end{equation}
where $\eps_1$ is the permittivity of the sphere embedded in a host medium of
permittivity $\eps_h$. This simple expression conveys a wealth of physical phenomena,
such as the position and the width of the localized surface plasmon (LSPR) or
plasmonic resonance ($\eps_1=-2\eps_h$). The very same expression can be found
in a large number of studies on the modeling of small resonant elements with dipole-like
radiation~\cite{DeVries1998,Kelly2003,Bricard2013,Manjavacas2014}.
Therefore, revisiting and refining the context of this expression and can potentially
have an effect on a wide range of disciplines.

In this work the concept of the polarizability is imposed to a more general case
of a sphere with a radially inhomogeneous (graded-index, or RI)
permittivity profile. This kind of profiles occur either naturally~\cite{Cai2017},
or as a result of sophisticated engineering processes~\cite{Zentgraf2011}.
However, this category can include even simpler structures, such as  core-shell
spheres, which are nothing but inhomogeneous spheres with step-wise permittivity
profile. Hence, by studying the properties of RI spheres we can expand our current
understanding on the effects of partially or continuously inhomogeneous scatterers.

The concept of an RI profile has a history. In optical and radio
engineering, the Luneburg, Eaton, and Maxwell fish-eye lenses are characteristic
examples of the utilization of inhomogeneity for tailoring the electromagnetic
scattering response~\cite{luneburg1964mathematical,Tai1958,Minano2006}. These first
examples where initially analyzed within the geometrical optics approximations.
Quickly after these problems where reformulated as a classical boundary value problem using Maxwell equations, and
rigorous remedies were available. Aden and Kerker, in their seminal work~\cite{Aden1951a}
delivered an \emph{\`a-la} Lorenz--Mie solution to the problem of a core-shell
structure, while Wait~\cite{Wait1962} generalized the step-wise homogeneous
problem for spheres with $n$-layers. In this way, any RI profile can be constructed
by a stratified sphere with variable permittivities for each layer. Undoubtedly,
this robust brute-force methodology solves the required scattering problem, however,
without offering physical intuition on the involved mechanisms.

An alternative treatment for the RI problem is the invariant imbedding
technique~\cite{Engheta1983,Johnson1988,Johnson1999}.~This technique
treats the step-homogenous problem in its infinite layer limit reformulating the
original scattering problem to a problem that satisfies a non-linear Riccati equation.
Subsequently, the resulting non-linear equation is numerically evaluated for a given
permittivity profile. Hence the invariant imbedding technique can be categorized
as a semi-analytical approach. Furthermore, the RI problem has been attacked by several
different volume-based numerical philosophies (see for example~\cite{Shalashov2016,Kolezas2017}).
~Apparently, all the aforementioned remedies are used mostly for designing the
scattering behavior of RI particles. The main drawback of these methods is that rarely
offer any physical insights on the involved scattering mechanisms and their particular characteristics.

One possible treatment that restores the physical intuition is obviously the exact solution
of the corresponding boundary value problem, and the formulated differential equation (second order)
of the radial function, for a given graded-index profile. A comprehensive overview
of the available exact solutions for the electrodynamic problem up to the
late 1960's can be found in~\cite{kerker2013scattering}, while more recent works
are listed in~\cite{babenko2003electromagnetic}. For instance, the exact solution
for a sphere with a Luneburg profile has been given by C.-T. Tai~\cite{Tai1959},
where the formulated radial differential equation is satisfied by a hypergeometric
function, while Westcott explored the available exact wave solutions for spherical
stratified media~\cite{Westcott1968}.

The corresponding problem of an RI subwavelength
sphere in the electrostatic (Rayleigh) limit and the available solutions of
inhomogeneous Laplace equation have been also reviewed in the past~\cite{Tai1963}. More recent works
on different RI profiles with power-law~\cite{Dong2003}, linear~\cite{Sihvola1988},
and polynomial profile (with some convergence restrictions)~\cite{Laquerbe2017}
brought into light some features of the deeply subwavelength RI problem;
a systematic study categorizing all the available analytical solutions for the electrostatic
problem is still missing.

\begin{figure}[!]
  \centering
  \includegraphics[width=0.475 \textwidth]{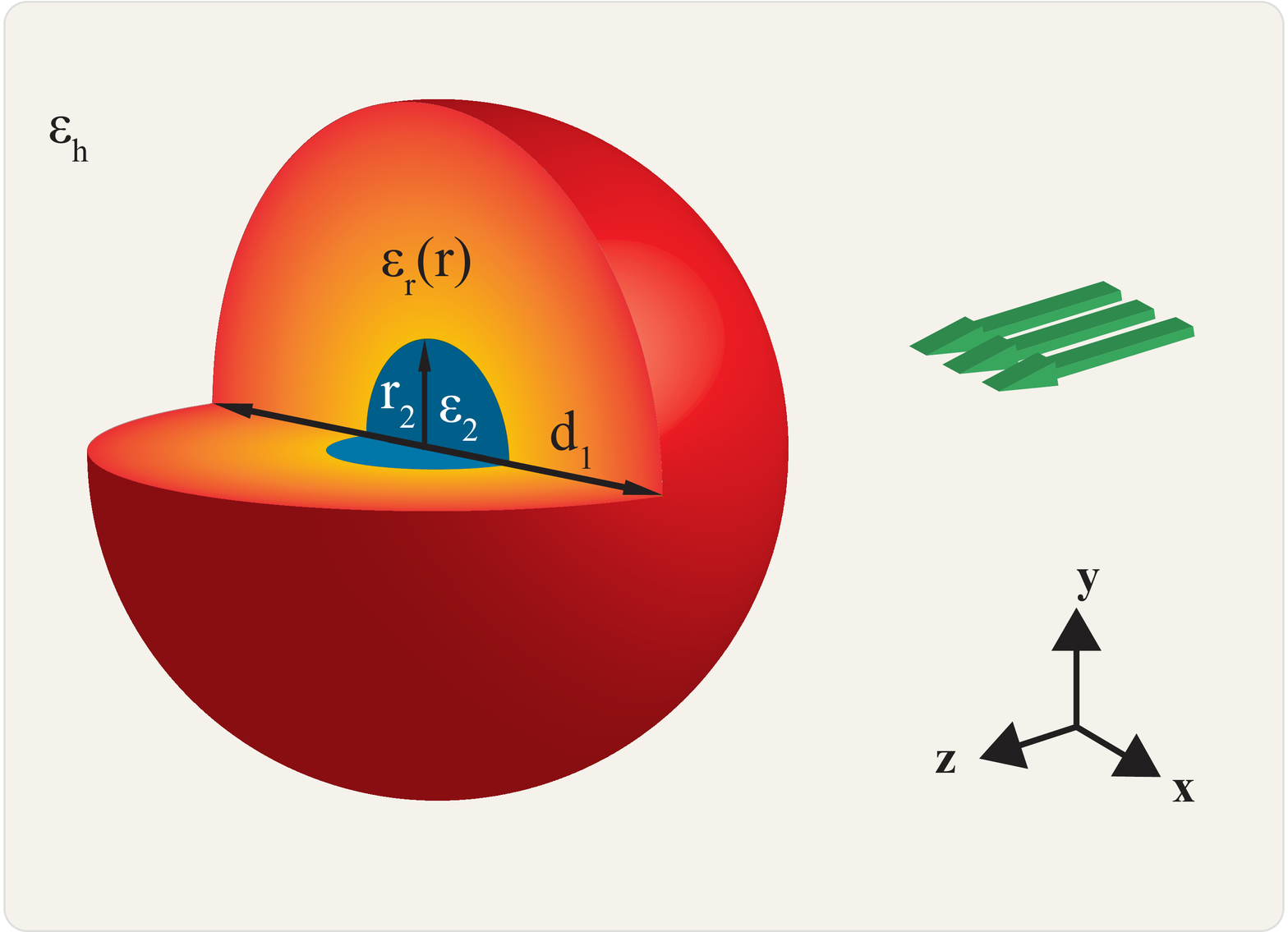}
\caption{A radially inhomogeneous sphere of diameter $d_1=2 r_1$ and radially inhomogeneous
permittivity $\eps_r(r)$, with an internal homogeneous core of radius $r_2$ and
permittivity $\eps_2$ immersed in a host medium ($\eps_h$) subject to a $z-$propagated
plane wave (constant excitation field in the long wave approximation).}
\label{fig:1}
\end{figure}

Inspired by the above developments and realizing the remaining gaps, we revisit the concept of polarizability towards a generalized
description that incorporates the effects of the inhomogeneous permittivity. Under this fresh perspective
we revisit the known case of a power-law profile and explore its scattering peculiarities.
The proposed model is further expanded towards a new family of permittivity profiles with an
exact solution, i.e, exponentially radial profiles.~Results on their resonant spectrum
reveal the existence of several scattering characteristics such as shifted
plasmonic resonances and peculiar scattering degeneracies for extreme permittivity values. The analysis
concludes with the validation of the presented results through a comparison
of the analytical scattering responses with that of a multilayer, step-homogeneous sphere,
fitted to the corresponding RI profiles.

The presented results can potentially stimulate further discussions about the
theory of RI scatterers, their particular functionalities, and open fertile grounds
toward their experimental implementation on modern energy control/harvesting applications.
The generalization of the polarizability can be particularly useful also in connection with inverse
scattering problems~\cite{colton2012inverse}, where the main objective often is the identification of unique material
inhomogeneities encoded in the signatures of the observed spectra.

\section{Theory}\label{sec:Theory}

Let us assume a sphere (Fig.~\ref{fig:1}, with subscript 0 for external, 1 for
shell, and 2 for core domain) of radii $r_1$ and $r_2$, subject to a uniform ($z-$directed)
electrostatic field causing a scattering potential of dipolar character, i.e.,
\begin{equation}
 \Phi_0(r,\theta)=\left( -E_0 r+\frac{B_0}{r^2}\right)\cos \theta
\end{equation}
Since the core region consists of a homogeneous material we have
\begin{equation}
 \Phi_2(r,\theta)=A_2r\cos\theta
\end{equation}
while the potential in the shell region can be written as
\begin{equation}
 \Phi_1(r,\theta)=f(r)\cos\theta
\end{equation}
assuming an arbitrary radial function $f(r)$. The expressions of the scattered (external, region 0) and the
internal (core, region 2) field ($\vec E=-\nabla \Phi$) are divergenceless, and satisfy the other requirements
of the corresponding physical problem, i.e., the scattered
field vanishes at large distances and has no singularities at the
origin~\cite{jackson1975electrodynamics}. In a similar manner we also require
divergenceless  electric flux density is at the inhomogeneous region (shell, region 1), viz.,
\begin{equation}\label{eq:D}
 \nabla\cdot\vec D_1=-\nabla\cdot\left(\eps_r(r)\nabla\Phi_1(r,\theta)\right)=0
\end{equation}
resulting to the following O.D.E
\begin{equation}\label{eq:ODE}
 f\DP(r)+\left(\frac{2}{r}+\frac{\eps_r\P(r)}{\eps_r(r)}\right)f\P(r)-\frac{2}{r^2}f(r)=0
\end{equation}

The main focus here is to study the cases where Eq.~(\ref{eq:ODE}) obtains a
closed form solution, hence extracting an analytical expression for the unknown
scattering amplitudes by solving the formulated boundary value problem.
For these cases one can express the radial function as
\begin{equation}
f(r)=A_1A(r)+B_1B(r)
\end{equation}
where $A(r)$ is a ``constant-like'' and $B(r)$ is a ``dipole-type'' solution of the
radial function. For instance, for the case of a homogeneous profile we have
$A(r)=r$ and $B(r)=\frac{1}{r^2}$, implying that $A(r)$ is well-behaving at
the origin and $B(r)$ contains a singularity. However, this is not always true,
since $A(r)$ and $B(r)$ can both exhibit a singular behavior, as it turns out to be for
the exponential permittivity profile. In that case special mathematical regularization
treatment is required, e.g., introduction of a singularity subtracting region at the center.

The extension of the analysis including higher-order multipoles can be done in a similar
manner as described in~\cite{jackson1975electrodynamics,Sihvola1988,Dong2003}.
In this work, we concentrate only on the main dipole contribution since in many
practical cases subwavelength spherical inclusions induce primarily a dipole field for a given
plane wave excitation. This is due to the symmetry of the excitation. Different
sources, i.e., with different symmetries such as dipoles or focused beams,
couple efficiently also with higher order modes. For these cases a higher order
analysis/treatment is required, a subject left for future investigations.

The electric fields for each domain are
\begin{equation}
 \begin{split}
  \vec E_0(r,\theta)= & E_0\vec u_z+\frac{B_0}{r^3}\left(2\cos\theta\vec u_r+\sin\theta\vec u_\theta\right) \\
  \vec E_1(r,\theta)=&-\left(A_1 A^\prime(r)+B_1 B^\prime(r)\right)\cos \theta \vec u_r\\
  &+\left(A_1 A(r)+B_1 B(r)\right)\frac{\sin \theta}{r} \vec u_\theta \\
  \vec E_2(r,\theta)= & -A_2 \vec u_z
 \end{split}
\end{equation}
where $\vec u_z=\cos\theta\vec u_r-\sin\theta\vec u_\theta$.
The unknown scattering amplitudes, $B_0, A_1, B_1$, and $A_2$, can be evaluated
by applying the continuity of the tangential electric field and normal flux
density components, expressed in potential form, i.e.,
\begin{equation}
\partial_\theta\Phi_j(r_j,\theta)=\partial_\theta\Phi_{j+1}(r_j,\theta)
\end{equation}
and
\begin{equation}
\varepsilon_j\partial_r\Phi_j(r,\theta)|_{r=r_{j+1}}=\varepsilon_{j+1}\partial_r\Phi_{j+1}(r,\theta)|_{r=r_{j+1}}
\end{equation}
where $j=0,1$. This set of four linear equations can be compactly expressed
in a matrix form $\vec A \bar X=\vec b$ for which the four unknowns are
$\bar X=\left(B_0~ A_1~ B_1~ A_2\right)^T$. The system matrices read

\begin{equation}\label{eq:matrixA}
\mathbf{A}=\left(
\begin{array}{cccc}
 -\frac{1}{r_1^2} & A(r_1) & B(r_1) & 0
 \\
 \frac{2 \varepsilon_0}{r_1^3} & \varepsilon_r(r_1)A^\prime(r_1) & \varepsilon_r(r_1)B^\prime(r_1) & 0
 \\
 0 & A(r_2) & B(r_2) & -r_2
 \\
 0 & \varepsilon_r(r_2)A^\prime(r_2) & \varepsilon_r(r_2)B^\prime(r_2) & -\varepsilon_2 \\
\end{array}
\right)
\end{equation}
and the excitation vector is
\begin{equation}
\mathbf{b}= -E_0
\left(
\begin{array}{c}
 r_1 \\
 \varepsilon_0 \\
 0 \\
 0 \\
\end{array}
\right)
\end{equation}

Note that primes denote the differentiation with respect to $r$. The determination
of the scattering amplitudes at each region is reduced to a brute force matrix
inversion, an algebraically laborious but rather straightforward task.

After the modularization of the solution we can shift our focus on the available
permittivity profiles. First, we consider the power-profile which has been studied in~\cite{Dong2003}.
Secondly, the formulated electrostatic
differential equation of Eq.~(\ref{eq:ODE}) can be solved for another family of
radially inhomogeneous permittivity profiles, i.e., exhibiting exponential radial
dependence of the type $e^{(n r)^p}$, with two parameters $n$ and $p$ ($ p$ is integer).
In this work we focus on two particular exponential profiles, i.e., the linear $e^{n r}$ ($p=1$)
(exp for short), and the inverse-linear $e^{\frac{1} {n r}}$ ($p=-1$) (inv-exp).
The term ``linear'' corresponds to the power of the exponent $ n r$, and correspondingly its inverse $\frac{1}{n r}$.

Exponential profiles can be applied for several purposes. For instance, the exponential profile can be
approximated as $e^{(n r)^p}\approx1+ (n r)^p$ when $n r\rightarrow0$ , and
$p\geq 0$. Assuming that the arbitrary constant $n$ can be of the form
$\frac{\omega_p^2}{\omega^2 b}$, where $b$ is a normalization length, the
exponential profile gives an approximation of a Drude-like model where the
plasma frequency exhibits a radial dependence. The above observation can be
generalized for every exponential profile, allowing the implementation of such profiles also modeling of realistic permittivity
distributions~\cite{lundqvist2013theory,Christensen2014}.

Similar forms of exponential permittivity profiles have been used as a
phenomenological description for the problem of solution-solvent electrostatic
interactions~\cite{Castner1988}. Here, we take the metamaterial/composite material
approach and analyze exponential profiles from the scattering perspective. It is projected that these kinds of
profiles can be used in an effective material description of artificially engineered composites
or used as a modeling fit for experimentally extracted scattering response, where the
triggered mechanisms require a permittivity description beyond the standard step-wise homogenous model.

\subsection{Power-law profile}\label{seq:power}
We start our analysis by assuming a power-law profile, i.e.,
\begin{equation}\label{eq:power}
\varepsilon_r(r)=\varepsilon_1 \left(\frac{r}{b}\right)^n
\end{equation}
where $b$ is a normalization factor with units of meter and $n$ is the
power factor $\left(n\in {\rm I\! R}\right)$. By inserting the above profile
to the general O.D.E of Eq.~(\ref{eq:ODE}) we obtain
 \begin{equation}
  f \DP(r)+\frac{1}{r}\left(n+2\right)f(r)\P-\frac{2}{r^2}f(r)=0
 \end{equation}
known as the non-homogeneous Euler--Cauchy differential equation which
has a solution of the form
\begin{equation}
 f(r)=A_{1} r^{p_{1}}+B_{1}r^{p_{2}}
\end{equation}
where
\begin{equation}\label{eq:power-factor}
p_{1,2}=-\frac{1}{2}\left(n+1\mp\sqrt{(n+1)^2+8}\right)
\end{equation}
are the power factors of the solution. These factors exhibit certain interesting properties. For example, $p_1$
and $p_2$ are always of different sign for any value of $n$, and
their product is constant, i.e.,
\begin{equation}
p_1p_2=-2
\end{equation}
The limiting cases are
$\lim_{n\to+\infty} p_1(n)=0$ and $\lim_{n\to-\infty} p_1(n)=+\infty$, while $p_1(0)=1$.

\subsection{Linear exponential profile $e^{n r}$}\label{sec:linear}

The first exponential profile under examination is the linear exponential profile, i.e.,
\begin{equation}\label{eq:exp}
 \varepsilon_r(r)=\varepsilon_1 e^{n \frac{r}{b}}
\end{equation}
where $n$ can be any arbitrary real parameter, and $b$ is a normalization radius with units [m].
For sake of simplicity $b$ is equal to the external radius $ r_1$ and will be omitted.

Following Eq.~(\ref{eq:ODE}), the formulated O.D.E is
 \begin{equation}
  f \DP(r)+\frac{1}{r}\left(n r+2\right)f\P(r)-\frac{2}{r^2}f(r)=0
 \end{equation}
and its solution, expressed in terms of the corresponding $A(r)$ and $B(r)$ functions, reads
\begin{equation}\label{eq:A_exp}
 A(r)=\frac{1}{n}\left(1-\frac{2}{n r}+\frac{2}{n^2 r^2}\right)
\end{equation}
and
\begin{equation}\label{eq:B_exp}
 B(r)=\frac{e^{-n r}}{r^2}
\end{equation}
where both $A(r)$ and $B(r)$ are singular at the origin. This is a rather counterintuitive
fact since the profile (\ref{eq:exp}) is smooth over the center of the sphere. In order to
sidestep the effects of this singularity, a singularity subtracting core is required.

\subsection{Inverse-linear exponential profile $e^{\frac{1} {nr}}$}

The second exponential case is the inverse-linear exponential profile (inv-exp) expressed as
\begin{equation}\label{eq:eps_inv-exp}
\varepsilon_r(r)=\varepsilon_1 e^{\frac{b} {nr}}
\end{equation}
with $n$ being an arbitrary constant, similar to the exponential profile, and $b$
is the radius normalization factor, similar to the previous cases (for simplicity we assume $b=1 $).
This leads to
\begin{equation}
  f \DP(r)+\frac{1}{nr^2}\left(2nr-1\right)f\P(r)-\frac{2}{r^2}f(r)=0
 \end{equation}
and the associated functions are
\begin{equation}\label{eq:inv-exp_A}
 A(r)=\frac{2nr-1}{2n-1}
\end{equation}
and
\begin{equation}\label{eq:inv-exp_B}
 B(r)=n\frac{2nr+1}{2n-1}e^{-\frac{1}{nr}}
\end{equation}

At this end, an important remark can be drawn. In the ordinary homogeneous cases and the power-law profiles
 $A(r)$ is a well-behaving function at the origin ($r=0$), while $B(r)$ contains
a singularity, e.g., in the power-law profile $r^{p_1}$ is smooth at
the origin ($p_1>0$ for every $n$), while $r^{p_2}$ is singular ($p_2<0$ for every $n$).
This observation, however, does not hold for every solvable case presented here.

For example, for the exp profile, both $A(r)$ and $B(r)$ (Eqs.~\ref{eq:A_exp} and~\ref{eq:B_exp})
are singular at the origin and a singularity extracting region at the origin (core) is therefore
necessary for calculating the polarizability of the core-shell inclusion;
the intact case can be reached by taking the limit of a vanishing core ($r_2\to0$).

On the other hand, for the inverse-exponential profile both $A(r)$ and $B(r)$ (Eqs.~\ref{eq:inv-exp_A} and~\ref{eq:inv-exp_B})
are smooth at the origin when $n\leq0$, which is another counterintuitive result.
In particular, when $n\rightarrow\pm\infty$ the inv-exp profile is smooth at the center
since $\lim_{n\to\pm\infty}\eps_r(r)\approx\eps_1$. However, for the case when $n\to0^-$
we have either an epsilon-near-zero (ENZ) profile at the center ($\lim_{r\to 0^-}\eps_r(r) \approx 0$),
or $\lim_{r\to0^+}\eps_r(r)\approx\infty$. The latter is a form of a PEC-like (perfect electric conductor)
behavior. Equivalently, the same limits can be reached when
$r\to0$ and $n<0$ and $r\to0$ and $n>0$, i.e., ENZ and PEC behavior, respectively.

In any of the above cases, these permittivity-induced peculiarities require the
existence of a singularity regularization core region. Therefore, the consideration of a general core-shell
setup is necessary for a regular solution of both exponential profiles.

\section{Analysis and Discussion}\label{sec:Results}

Once the matrices $\vec A$, and $\vec b$ are assembled (see Section~\ref{sec:Theory}),
all the unknown fields amplitudes can be determined by the expression
$\bar X=\vec A^{-1}\cdot\vec b$. The external amplitude $B_0$ represents the amplitude
of the dipolar field created outside the scatterer,
quantifying the dipole strength caused by the presence of an external field.
In other words, parameter $B_0$ is nothing but the polarizability of the studied
inclusion~\cite{sihvola1999electromagnetic}.
The resulting static polarizability can be computed as follows
\begin{equation}
 \alpha_\text{s}= 3\frac{B_0}{r_1^3E_0}
\end{equation}

The extracted electrostatic polarizability quantifies only the electrostatic effects.
This model successfully captures the radiation enhancement (or Fr\"ohlich condition)
without, however, taking into account any kind of radiation reaction that restores
the conservation of energy for this physical system~\cite{DeVries1998,Carminati2006,Moroz2010}.
To do so, an imaginary term accounting for the radiation reaction needs to be introduced
to restore energy balance of this passive system. This new corrected (or Modified
Long Wave Approximation-MLWA~\cite{Moroz2010})
quasistatic polarizability reads~\cite{Tretyakov2014,Tzarouchis2016c}
\begin{equation}\label{eq:a_dyn}
\alpha_d=\frac{-i\frac{2}{9}\alpha_{s}x^3}{1-i\frac{2}{9}\alpha_{s}x^3}
\end{equation}
where $x=kr_1$ is the size parameter relative to the host medium ($k$ is the
wavenumber of the host medium). Consequently, the scattering and extinction
efficiencies are written
\begin{equation}\label{eq:Q_s}
Q_{sca}=\frac{6}{x^2}|\alpha_d|^2
\end{equation}
In the following sections all the scattering and extinction efficiencies
depicted are given by Eq.~(\ref{eq:Q_s}), accounting also for the radiative
reaction effects. In this sense the validity of the proposed model expands
beyond the electrostatic regime, and can be readily used up to moderate sized spheres.

At this point it is necessary to refine the way for expressing the extracted
dipole scattering amplitude (polarizability). The inversion of matrix~(\ref{eq:matrixA}) for a
general $\eps_r(r)$ profile can be compactly written as
\begin{equation}\label{eq:general}
 B^\text{general}_0=\frac{C\eps_r(r_1)-\varepsilon_0}{C\eps_r(r_1)+2\varepsilon_0}E_0r_1^3
\end{equation}
where
\begin{widetext}
  \begin{equation}\label{eq:general_C}
 C=r_1\frac{\varepsilon_2\left(A^\prime(r_1)B(r_2)-A(r_2)B^\prime(r_1)\right)+\varepsilon_r(r_2)r_2\left(A^\prime(r_2)B^\prime(r_1)-A^\prime(r_1)B^\prime(r_2)\right)}{\varepsilon_2\left(A(r_1)B(r_2)-A(r_2)B(r_1)\right)+\varepsilon_r(r_2)r_2\left(A^\prime(r_2)B(r_1)-A(r_1)B^\prime(r_2)\right)}
  \end{equation}
\end{widetext}
is the inhomogeneity factor. This complicated but rather straightforward expression
conveys elegantly the effects of the inhomogeneity in a core-shell structure.

The form of Eq.~(\ref{eq:general}) facilitates the analysis of profiles
that do not have an analytical solution, expanding its significance for inverse-scattering problems.
For a given (arbitrary) material profile one can either experimentally
or numerically extract the scattering spectrum
of a given inhomogeneous sphere, and fit the phenomenological description of Eq.~(\ref{eq:general_C}) to
the observed spectrum by properly adjusting the inhomogeneity factor $C$. Therefore, the inhomogeneity
factor is directly applicable for inverse-engineering/scattering purposes.

An intact sphere can be approached by taking the limit of the inhomogeneity
factor $C$ when~$r_2\to0$ in Eq.~(\ref{eq:general_C}). Generally, this is a function of both
$A(r)$ and $B(r)$. However, for the standard case of
well-behaved functions, i.e., $\lim_{r\to0}A(r)=0$ and $ \lim_{r\to0}B(r)=\infty$,
the inhomogeneity factor reduces into a function of the $A(r)$ and its value on the
external boundary of the sphere, viz.,
\begin{equation}
 C=r_1\frac{A^\prime(r_1)}{A(r_1)}=r_1\left[\ln(A(r))\right]^\prime|_{r=r_1}
\end{equation}

The compact form of polarizability in Eq.~(\ref{eq:general}) might be of particular
use also for the plasmonic scattering enhancement case, since it generalizes the
resonance condition in a simple manner, i.e.,
\begin{equation}
 \varepsilon_r(r_1)=-\frac{2\varepsilon_0}{C}
\end{equation}
One can observe that the inhomogeneity factor contributes directly to the main
resonant condition. Finally, when the radiative reaction is included (Eq.~(\ref{eq:a_dyn}))
the resonance leads to the following complex expression
\begin{equation}
  \varepsilon_r(r_1)=-\frac{2\varepsilon_0}{C}-i\frac{2\varepsilon_0}{C}x^3
\end{equation}
implying that the inhomogeneity factor modifies not only the resonant position
but also the width and the maximum resonant
absorption of plasmonic resonance~\cite{Tretyakov2014,Tzarouchis2016c}.

Exact forms of the extended quasistatic polarizability
is of paramount importance for applications focused in controlling light-matter interactions,
either as single scattering effects or as collective effects in composite devices.
For instance, the dispersion engineering problem can be approached by analytic, closed-form polarizability
expressions; the effective medium description is proportional
to the single inclusion  polarizability~\cite{sihvola1999electromagnetic,Kim2018}.

Additionally, knowledge of the RI polarizability
gives direct inspection for the limiting behavior of multilayer particles, especially
when the number of layers increases.  This limiting process obviously can
lead to simpler and physically intuitive understanding regarding which parameters
are affecting the scattering behavior of the proposed scatterer.

Actually, there is a remarkably large amount of modern applications, from
 optical antennas~\cite{Bharadwaj2009} to plasmonic devices and sensors~\cite{Kelly2003},
optical forces~\cite{Gao2015}, and hot-electron photocalaysis~\cite{Manjavacas2014,Brongersma2015}
where polarizability expressions give direct estimations on the scattering and absorptive
characteristics of the corresponding systems. In all the above cases, engineered or stochastic (random) inhomogeneities
of a single particle may significantly alter the overall scattering behavior.
Therefore, closed-form solutions of RI particles allow the direct study and implementation of these phenomenas.

\subsection{Homogeneous Core-Shell: Inhomogeneity perspective}

A core-shell particle with homogeneous regions (core $\varepsilon_2$, and shell,
 $\varepsilon_1$) is perhaps the simplest case of an inhomogeneous profile, with
  a step-wise function character. For this case ($A(r)=r$ and $B(r)=\frac{1}{r^2}$)
   the polarizability reads
\begin{equation}
 B^\text{cs}_0=\frac{C\varepsilon_1-\varepsilon_0}{C\varepsilon_1+2\varepsilon_0}E_0r_1^3
\end{equation}
where the introduced coefficient is
\begin{equation}\label{eq:c_core-shell}
C_\text{cs}=-2+3\frac{1}{1-\frac{\varepsilon_2-\varepsilon_1}{\varepsilon_2+2\varepsilon_1}\eta^3}
\end{equation}
and $\eta=\frac{r_2}{r_1}$ is the radius ratio. The scaling factor $C$ approaches
 unity when either $\eta=0$ (no core) or the contrast between the core and shell
  permittivity is zero ($\varepsilon_1-\varepsilon_2=0$). On the other hand
  $C=\frac{\eps_2}{\eps_1}$ for $\eta\to1$, and the expression leads to the well-known
  polarizability for a core-shell sphere~\cite{Tzarouchis2018a}

\begin{equation}\label{eq:core-shell}
 B^\text{cs}_0=\frac{(\varepsilon_1-\varepsilon_0) (\varepsilon_2+2\varepsilon_1)+\eta ^3 (\varepsilon_0+2 \varepsilon_1) (\varepsilon_2-\varepsilon_1)}{(\varepsilon_1+2\varepsilon_0) (\varepsilon_2+2\varepsilon_1)+2 \eta ^3 (\varepsilon_1-\varepsilon_0) (\varepsilon_2-\varepsilon_1)}E_0 r_1^3
\end{equation}

\subsection{Power profile}

The polarizability of an intact sphere with a power profile is one of the most studied profiles~\cite{Dong2003,Reshetnyak2016}.
Although many aspect of such sphere are known, here we focus on the aspects that
have not previously discussed. In this direction, the generalized polarizability
of a core-shell inhomogeneous sphere with a power profile (Eq.~(\ref{eq:power})) reads

\begin{equation}
B_0^\text{power}=\frac{C\eps_r(r_1)-\eps_h}{C\eps_r(r_1)+2\eps_h}
\end{equation}
where the inhomogeneity factor is
\begin{equation}\label{eq:power-2}
C_\text{power}=p_2+\frac{p_1-p_2}{1-\frac{\eps_2-p_1\eps_r(r_2)}{\eps_2-p_2\eps_r(r_2)}\eta^{p_1-p_2}}
\end{equation}
Apparently, this result reduces to Eqs.~(\ref{eq:c_core-shell}) and~(\ref{eq:core-shell})
when $n=0$, $p_1=1$, $p_1=-2$, $\varepsilon_r(r_1)=\varepsilon_1$,
and $\varepsilon_r(r_2)=\varepsilon_1$.

\begin{figure}[!]
\centering
\includegraphics[width=1 \textwidth]{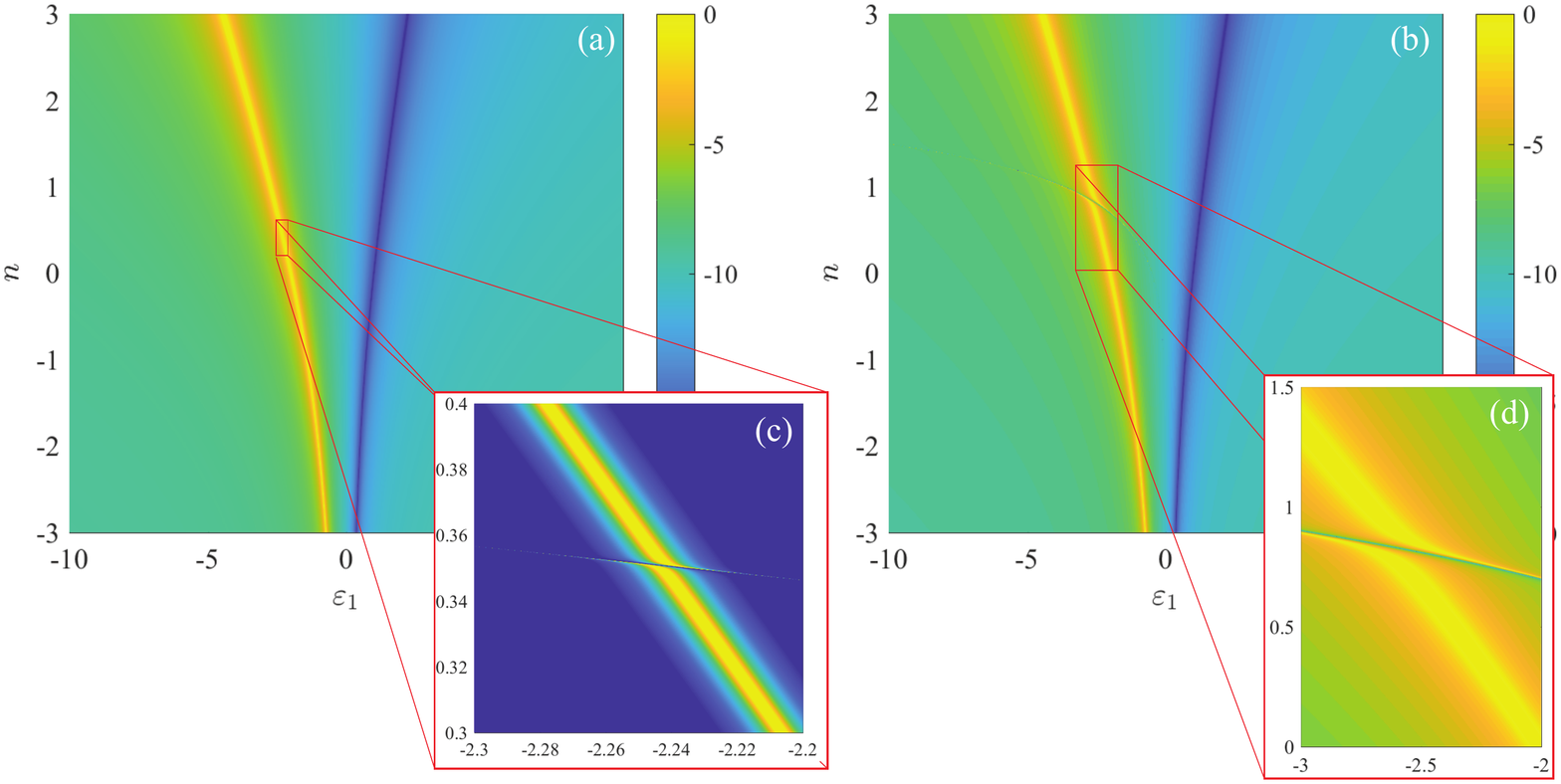}
\caption{The polarizability as a function of the power factor $n$, and permittivity
 $\eps_1$ for (a) $\eta=0.01$ (very small core radius) and (b) $\eta=0.1$.
 The inset figures (c) and (b) depict the existence of a pole-zero scattering
 crossing for both cases. Note that in figure (c) the colorscale is adjusted
 for better visualization of the crossing.}
\label{fig:power}
\end{figure}

One interesting feature is the radius ratio dependence in  Eq.~(\ref{eq:power-2})
where a power law has an exponent of $p_1-p_2=\sqrt{\left(n+1\right)^2+8} $.
This interesting feature reveals that a sphere with a power-law profile exhibits
plasmonic resonances with different ``fraction volume'', $\eta^{p_1-p_2}$, than
the volume dependence ($\eta^3$) observed in a homogeneous core-shell case~\cite{Tzarouchis2017e}.
Note that for $n=-1$ this exponent obtains its global minimum value, $2\sqrt{2}$.

The simpler case of an intact inhomogeneous sphere, i.e., $\eta\rightarrow0$
results in
\begin{equation}\label{eq:solid-power}
 B_0=\frac{p_1\varepsilon_1(r_1)-\varepsilon_0}{p_1\varepsilon_1(r_1)+2\varepsilon_0}
\end{equation}
since the inhomogeneity parameter is $C=p_1$. As can be seen, Eq.~(\ref{eq:solid-power})
exhibits resonant behavior when the condition
$\varepsilon_r(r_1)=-\frac{2}{p_1}\varepsilon_0$ is satisfied. For the limiting cases
when $n\rightarrow+\infty$ the power factor $p_1\rightarrow0$ implying that
$\varepsilon(r_1)\rightarrow-\infty$, while for $n\rightarrow-\infty$ we have
$p_1\rightarrow+\infty$ and $\varepsilon(r_1)\rightarrow0$. Similar
trends can be deduced for the scattering minimum. Figures~\ref{fig:power} (a)
and (b) give the scattering efficiency as a function of the power factor $n$
and the internal permittivity $\eps_1$ for the cases where $\eta=0.01$
and $0.1$, Fig.~\ref{fig:power} (a) and Fig.~\ref{fig:power} (b), respectively.

It is interesting to notice that the existence of a core gives a zero-pole
scattering crossing for values close to $n=0.36$ (Fig.~\ref{fig:power} (c)) and
$n=0.86$ (Fig.~\ref{fig:power} (d)). This scattering crossing can be recognized
as an embedded eigenmode, a scattering degeneracy with extremely sharp
characteristics, able to support eigenmodes with infinite lifetimes for the
lossless case~\cite{Silveirinha2014,Monticone2014}. This
type of degeneracy is attributed solely to the existence of a core to the system.
At the vicinity of the centre of the sphere the inhomogeneous permittivity shell exhibits an
ENZ behavior allowing the extreme confinement of electric field
(see for example the review on the properties of ENZ structures~\cite{Liberal2017}).
%

\subsection{Exponential profiles}

The analysis of the previous section can be repeated in a rigorous manner for the less-studied exponential
profiles. In an attempt to deliver some first insights regarding the scattering
peculiarities of such sphere we analyze the case of an intact sphere.
As can be seen already from their mathematical treatment in Section~\ref{sec:Theory}, these particular
profiles require the existence of a regularization core region. Therefore, the case
an intact sphere and its polarizability can be approached by repeating the analysis
of a core-shell structure and taking the limit case of a vanishingly small core ($\eta\rightarrow0$).

\begin{figure}[!]
\centering
\includegraphics[width=1 \textwidth]{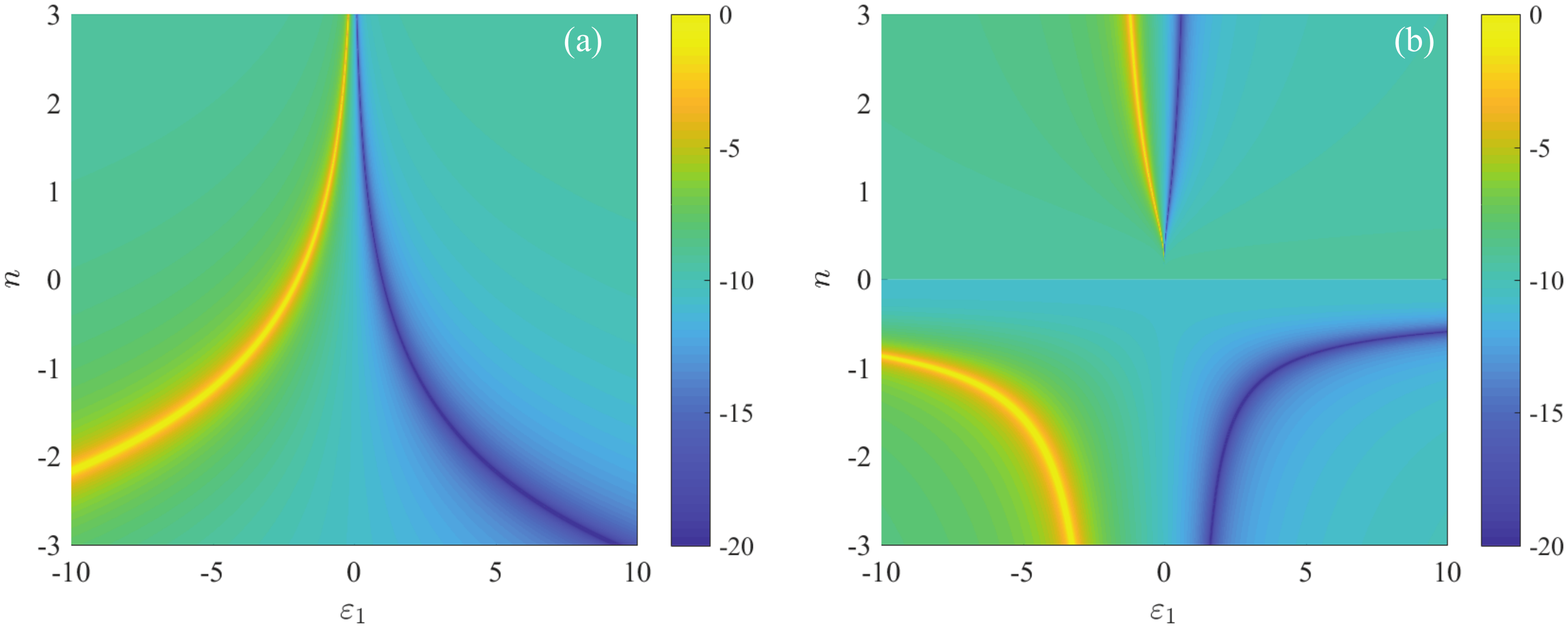}
\caption{Scattering efficiency (in logarithmic scale) of (a) the linear exponential
and (b) the inverse-linear exponential permittivity profiles as a function
of the scaling factor $n$ and the internal permittivity $\eps_1$. The yellow regions correspond
to the scattering enhancement and the blue regions to scattering minimum. The colorscale has been adjusted
for better visualization of the results.}
\label{fig:exp}
\end{figure}

Starting with the linear exp profile, $e^{nr}$, the inhomogeneity factor of an
intact sphere ($\eta\to0$) reduces to
\begin{equation}
 C_\text{exp}=2\frac{e^{n}\left(n -2\right)+n+2}{e^{n}\left(n^2 -2 n +2\right)-2}
\end{equation}
where for $n\rightarrow0$ we have $C_\text{exp}=1$. Figure~\ref{fig:exp} (a)
depicts the plasmonic resonance spectrum at the parametric space of $\eps_1$ and the factor $n$.
One observes that in this case, large positive $n$ values shift both the scattering
zero (Fig.~\ref{fig:exp} (a) blue valley) and the pole (Fig.~\ref{fig:exp} (b) yellow peak) toward the ENZ region.
At the limit where $n\to +\infty$ both features (zero-pole) finally cancel each other, creating
an ultra-sharp resonant degenerative point, a form of an embedded eigenstate~\cite{Monticone2014}.
However, negative $n$ values lead to a strong shifts of both features towards opposite directions.
Obviously when $n=0$, the pole and the zero coincide with the homogeneous case,
i.e., $\eps_1=-2$ and $\eps_1=1$, respectively.

In a similar manner, the inverse linear exponential profile gives an inhomogeneity factor of the form
\begin{equation}
 C_\text{inv-exp}=
\begin{cases}
1+\frac{1}{n}-\frac{1}{2n+1} & \text{, } n>0\\
\frac{2n}{2n-1} & \text{, } n\leq0
\end{cases}
\end{equation}
In this case the distribution of the inhomogeneity factor $C_\text{inv-exp}$
depends on the sign of the parameter $n$. The corresponding spectrum of the inv-exp case (Fig~\ref{fig:exp} (b))
reveals an reverse trend with respect to the exp case. The homogeneous-profile case with the resonance close to $\eps_1=-2$ and
scattering minimum at $\eps_1=1$, can be asymptotically reached for $n\to\pm\infty$, as can be seen in  Fig~\ref{fig:exp} (b).
On the other hand, small negative values of $n$ cause an extreme shift of the observed resonance, while
the case of small positive values of $n$ result in the collapse of both pole and
zero at the ENZ limit, like in the limit $n\to+\infty$ in the exp case.

\subsection{Validating the results}

Finally, in order to further verify our results, a comparison with a step-homogeneous
multilayered case sphere is implemented. The polarizability of a multilayered sphere
can be enumerated following an iterative analysis presented in~\cite{Sihvola1988}.
The permittivity of each layer exhibits a constant value, extracted from the continuous profile.
In this sense, the permittivity profile of the multilayered sphere is nothing but
the discretized version of each of the aforementioned continuous permittivity profiles.

\begin{figure}[!]
\centering
\includegraphics[width=1 \textwidth]{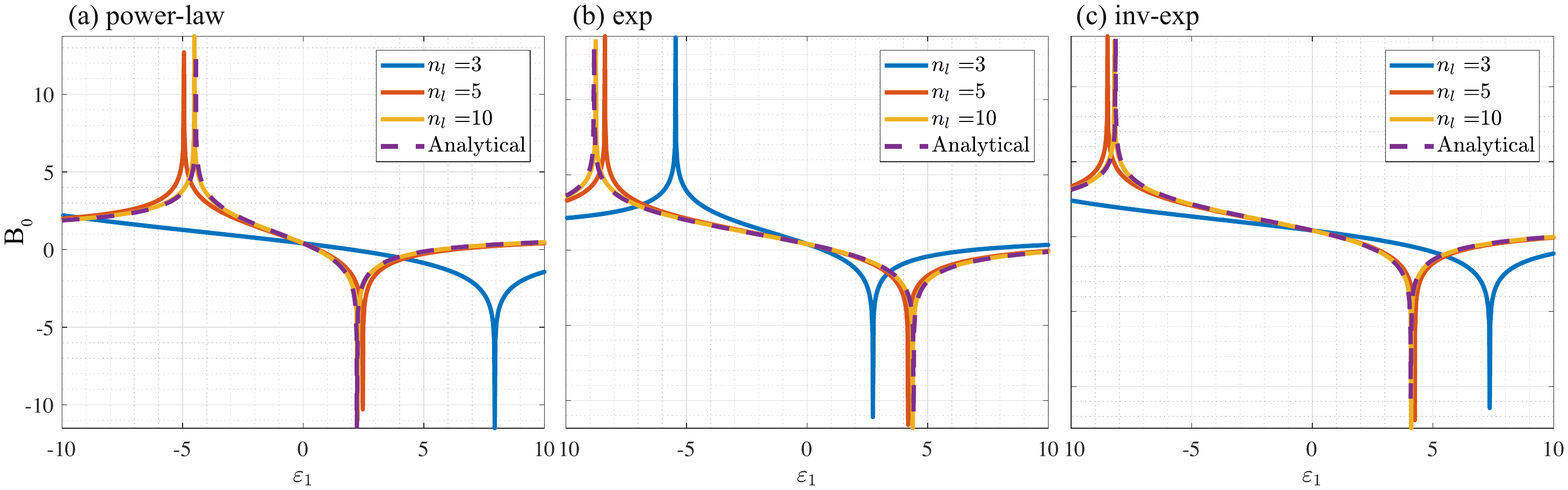}~
\caption{The scattering amplitude $B_0$ (colors in adjusted logarithmic scale) for all three
analytical cases (dashed purple lines) compared with the response of a
multilayered sphere of $n_l=3$ (blue lines), $n_l=5$ (red lines), and $n_l=10$
(orange lines) layers. Subfigures depict (a) the power profile for $n=2$, (b)
the exponential profile for $n=-2$, and (c) the inv-exp profile for $n=-1$.
A multilayer sphere with more than $n_l=10$ can accurately capture the scattering
trends, verifying in this way the validity of the theoretical analysis}
\label{fig:multi_all}
\end{figure}

Figure~\ref{fig:multi_all} depicts a comparison for the three cases, as a function
of the number of layers. As we can see, a multilayered structure with more than
$n_l=10$ layers reproduces the scattering behavior very accurately.
In the computations, the permittivity of layers is taken as the geometrical mean value between
the external and internal radius of each layer.

\section{Conclusions}

The expression of Eq.~(\ref{eq:general}) generalized the concept of the
homogeneous polarizability, allowing us to rigorously explore the non-trivial
physical mechanisms for a whole new family of graded-index particles. The introduced
polarizability description offers a direct homogenization formula for the effective permittivity
perspectives in both realistic and engineered manner ~\cite{Chettiar2012}. Additionally, this description
can be used for reverse-engineering the inhomogeneity coefficient $C$, fitting to experimental or numerical data.
In this way the experimentally observed deviations of the
plasmonic resonances on deeply subwavelength spheres~\cite{Christensen2014} can potentially find
simpler phenomenological interpretation; the same general form of polarizability can be extrapolated even for
non-analytically solvable profiles, allowing an approximative estimation regarding their behavior.

Aside from the above cases, the analytical study can be implemented for a wide range of
applications where rigorous modeling of artificially grown inhomogeneous structures, e.g., stratified
spheres, transformation optics~\cite{Vakil2011}, irregularly-shaped particles~\cite{Monreal2017},
is required. In simple words, the proposed inhomogeneously-refined
model can directly replace the widely-used homogeneous polarizability description for each of the aforementioned cases.

The same analysis can be also deployed for thermo-plasmonic applications where the existence of
a temperature gradient changes the permittivity distribution, causing an effective inhomogeneous profile.
These thermally-induced inhomogeneities might lead to improved heat-assisted magnetic recording~\cite{Boriskina2017}
and reinforce our deeper understanding on heat diffusion problems in plasmonic nanopartilces~\cite{LukYanchuk2012}.

Lastly, diffusive effects, especially between interfaces that follow radially dependent distribution
can be easily approached by the introduced models, for example in heavily doped semiconductor scatterers
offering a plethora of new absorption and scattering functionalities~~\cite{Khurgin2018}.
In conclusion, it is envisioned that the presented study will stimulate novel energy control/harvesting
ideas for nanophotonic applications, such as the implementation of subwavelength
plasmonic particles exhibiting Luneburg, Eaton, or more exotic graded-index profiles.

\section*{Acknowledgements}
The work is supported by the Aalto University ELEC Doctoral School Scholarship.
D.C.T would like to thank prof. Nader Engheta for the host at University of Pennsylvania,
where parts of this work have been completed.
\bibliography{article_main}

\end{document}